\documentclass{moriond}

\def\be{\begin{equation}}
\def\ee{\end{equation}}
\def\bea{\begin{eqnarray}}
\def\eea{\end{eqnarray}}

\newcommand{\Photo}{\includegraphics[height=35mm]{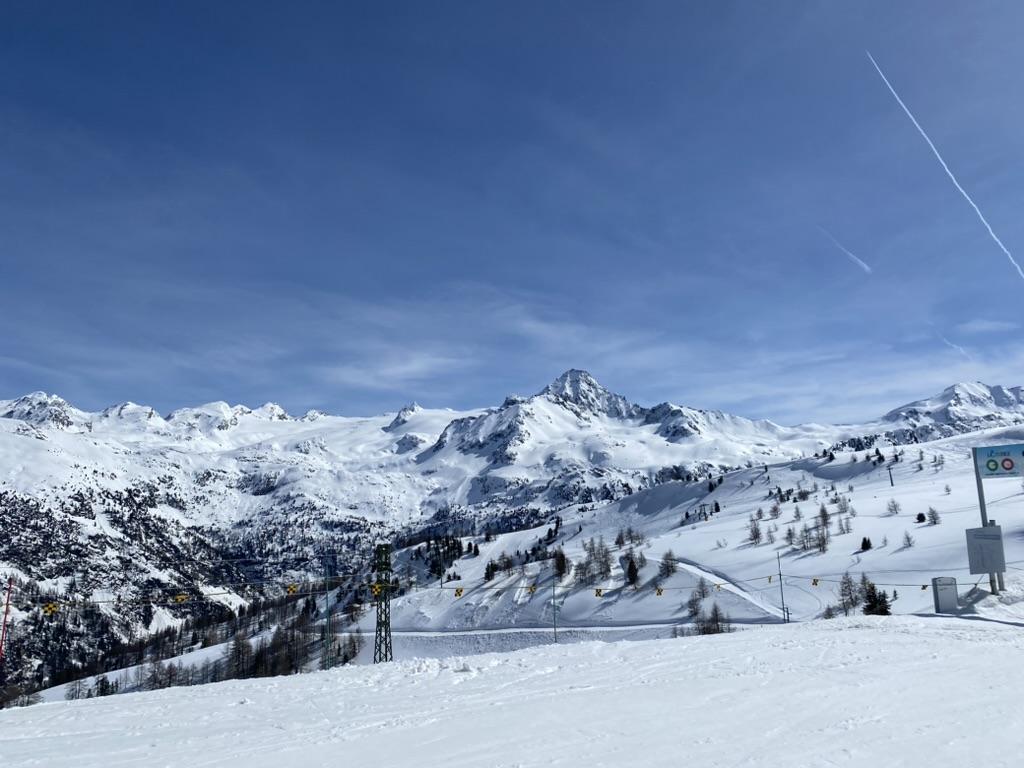}}

\begin{document}
\vspace*{4cm}
\title{Recent Cold QCD Results from STAR}

\author{ Ting Lin for the STAR Collaboration }

\address{Institute of Frontier and Interdisciplinary Science \& Key Laboratory of Particle Physics and Particle Irradiation (MoE), Shandong University, Qingdao, Shandong 266237, China}

\maketitle\abstracts{Understanding the origin of transverse single-spin asymmetries is a long-standing challenge in strong interaction physics. Recent precise measurements of the azimuthal distribution of charged pions in jets from STAR have shed new light on the spin–momentum correlations within the Transverse Momentum Dependent (TMD) formalism. This measurement, particularly sensitive to the Collins effect, correlates the quark transversity with the Collins fragmentation functions, elucidating the spin-dependent fragmentation process. Moreover, similar measurements involving kaons and protons, which have unique quark compositions, provide vital insights into the flavor-dependent spin dynamics of the parton distribution functions and fragmentation functions. These studies are crucial for developing a more comprehensive understanding of nucleon spin structure. In these proceedings, selection of the recent Cold QCD STAR results are presented, focusing on the asymmetries for charged kaons and protons in jets from transversely polarized $pp$ data at $\sqrt{s}$ = 200 GeV.}

\section{Introduction}
\label{intro}
The Transverse Momentum Dependent (TMD) formalism elucidates the complex spin–momentum correlations between nucleon and partons inside it, providing critical insights into the three-dimensional structure of hadrons in quantum chromodynamics (QCD). In the TMD physics, initial state interactions are characterized by eight independent TMD parton distribution functions (PDFs). They describe the distribution of partons carrying a longitudinal momentum fraction $x$ and transverse momentum $k_{T}$ within an initial state hadron in different spin states. Similarly, the dynamics of final state interactions are captured by eight TMD fragmentation functions (FFs), which describe the hadronization of a parton into a final state hadron which carries a longitudinal momentum fraction $z$ and a transverse momentum $\kappa_T$ relative to the fragmenting quark~\cite{Boussarie:2023izj}. The Collins effect~\cite{Collins:1993} highlights these final-state spin-momentum correlations, illustrating how a transversely polarized quark fragments into an unpolarized hadron, with a preference for the hadron to be emitted sideways relative to the quark's spin axis. It sheds light on the hadronization process, demonstrating how the spin of quarks influences the angular distribution of the emitted hadrons.\par

\begin{figure*}
\centering
\includegraphics[width=0.65\linewidth]{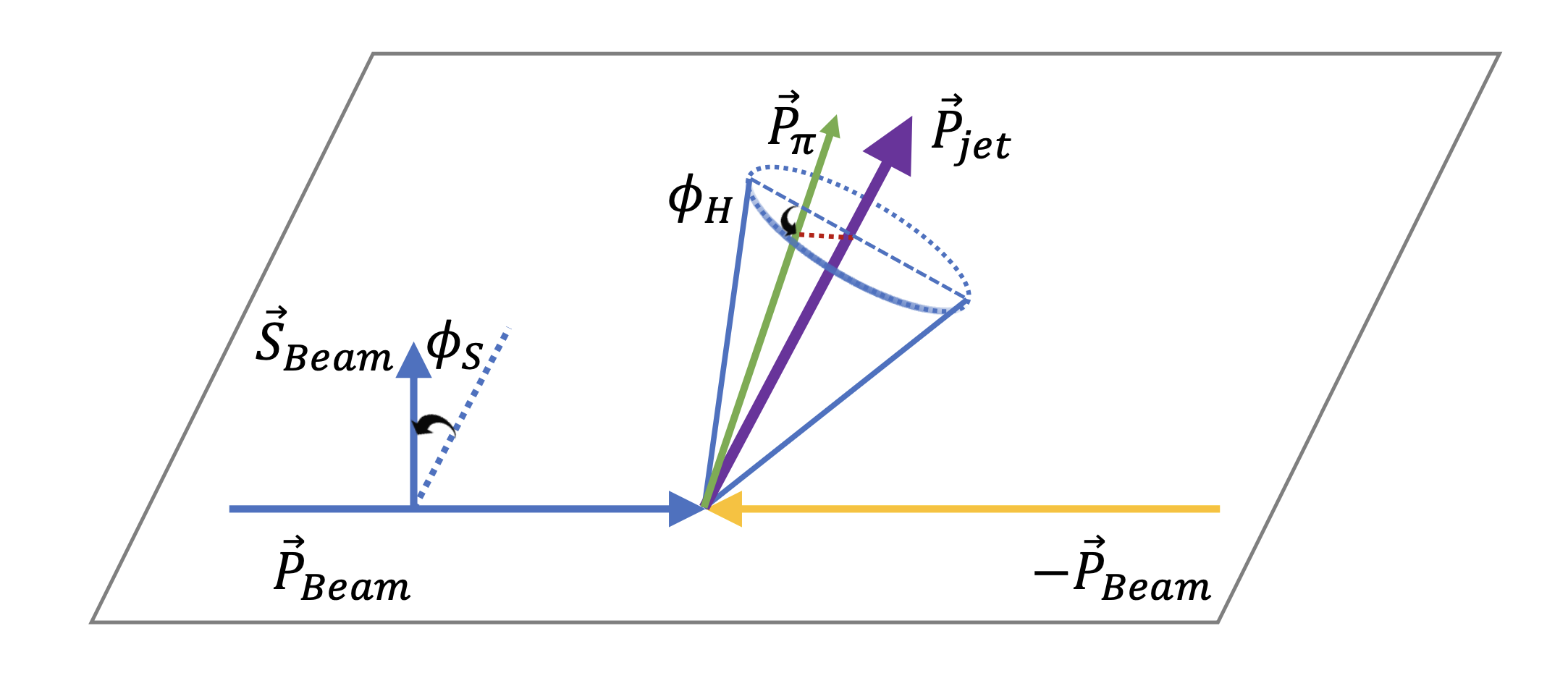}
\caption{Definition of azimuthal angles $\phi_{S}$ and $\phi_{H}$ in polarized hadronic collisions.}
\label{fig:azimuthal_modulation}       
\end{figure*}

The Relativistic Heavy Ion Collider (RHIC) stands as a unique facility in the exploration of the TMD physics with its ability to accelerate and collide transversely polarized proton beams at center-of-mass energies of 200 and 510~GeV~\cite{Alekseev:2003sk}. In transversely polarized proton-proton collisions, the Collins effect can be probed through hadron-in-jets production. In this channel, an asymmetry arises through the correlation of the transverse spin of a fragmenting quark with the transverse momentum of the final state hadron relative to the jet axis~\cite{Yuan:2007nd}~\cite{Alesio:2011PRD}~\cite{Alesio:2017PLB}~\cite{Kang:2017JHEP}~\cite{Kang:2017PLB}. This asymmetry can be further expressed as:
\begin{equation}
\label{equ:spin_cross_section_AN}
A_{UT}\sin(\phi_{S} - \phi_{H}) = \frac{\sigma^{\uparrow}(\phi_{S},\phi_{H}) - \sigma^{\downarrow}(\phi_{S},\phi_{H})}{\sigma^{\uparrow}(\phi_{S},\phi_{H}) + \sigma^{\downarrow}(\phi_{S},\phi_{H})} 
\rightarrow
\frac{\sum_{abc}h_{1}^{a}(x_{1},\mu)f_{b}(x_{2},\mu)\sigma^{\mathrm{Collins}}_{ab\rightarrow c}H_{1,h/c}^{\perp}(z_{h},j_{T};Q)}{\sum_{abc}f_{a}(x_{1},\mu)f_{b}(x_{2},\mu)\sigma^{\mathrm{unpol}}_{ab\rightarrow c}D_{h/c}(z_{h},j_{T};Q)}
\end{equation}
where $h_{1}^{a}(x_{1},\mu)$ is the quark collinear transversity, while $H_{1,h/c}^{\perp}(z_{h},j_{T};Q)$ is the TMD Collins fragmentation function. $f_{b}(x_{2},\mu)$ and $D_{h/c}(z_{h},j_{T};Q)$ are the unpolarized parton distribution function and fragmentation function. $\sigma^{\mathrm{unpol}}$ is the unpolarized partonic cross section while $\sigma^{\mathrm{Collins}}$ is the spin-dependent partonic cross section. Here $\phi_{S}$ represents the azimuthal angle between the polarization of the proton beam to the jet scattering plane, and $\phi_{H}$ is the azimuthal angle of the hadrons inside the jet relative to the jet scattering plane as shown in Fig.~\ref{fig:azimuthal_modulation}.\par

\begin{figure*}
\centering
\includegraphics[width=0.65\linewidth]{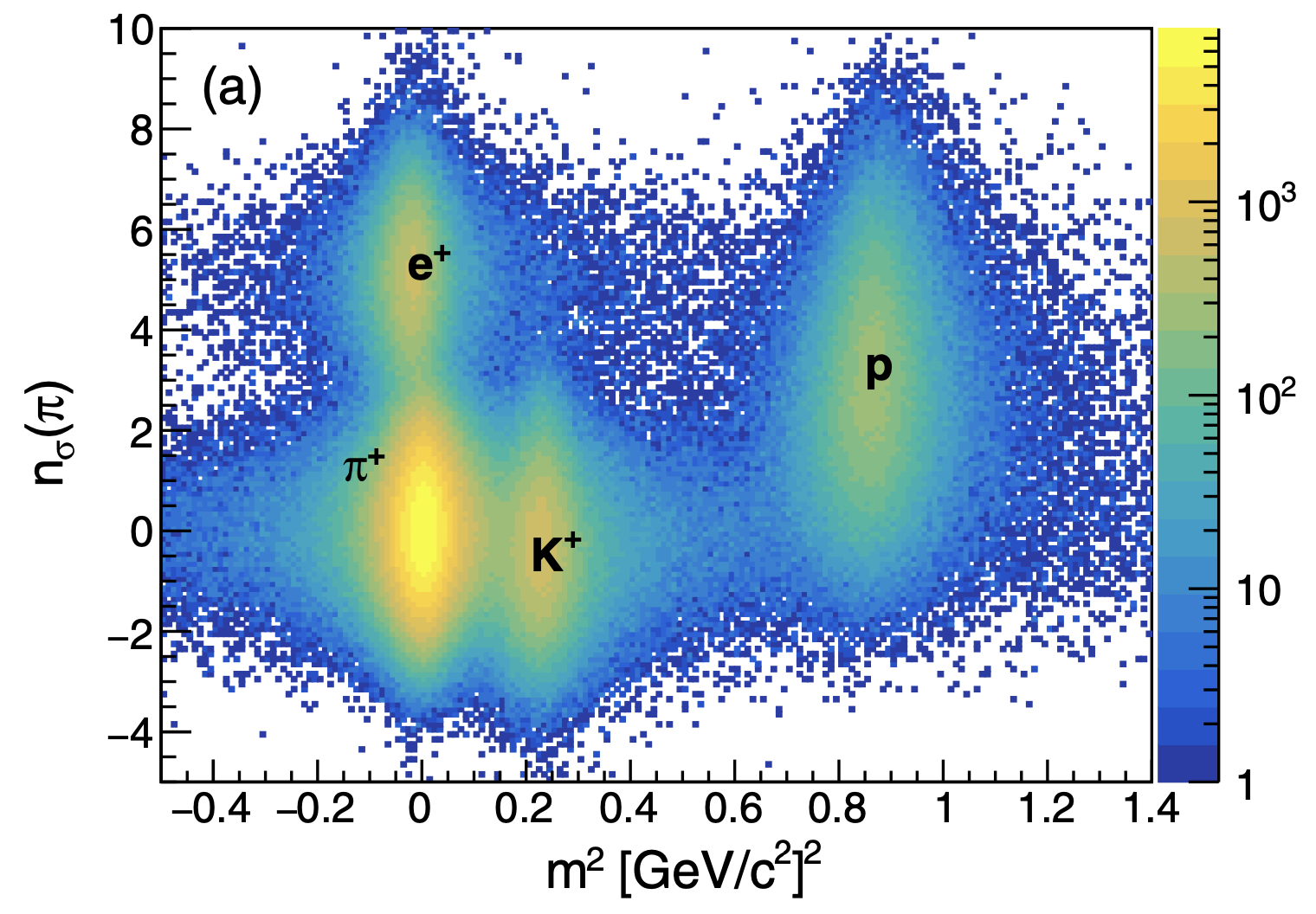}
\caption{The correlations of $n_\sigma(\pi)$ \textit{vs}.\@ $m^{2}$ for positively charged particles carrying momentum fractions of 0.1 $< z <$ 0.13 in jets with 8.4 $< p_{T} <$ 9.9 GeV$/c$.}
\label{fig:pid}       
\end{figure*}

The Solenoidal Tracker at RHIC (STAR), with its wide kinematic coverage and particle identification capability, provides a good opportunity for this measurement~\cite{STAR:2002eio}. The STAR Time Projection Chamber (TPC) reconstructs the charged particles with high precision~\cite{Anderson:2003ur}, while the electrons and photons are efficiently detected by the Barrel and Endcap Electromagnetic Calorimeter (BEMC and EEMC)~\cite{Beddo:2002zx}~\cite{Allgower:2002zy}. To identify specific charged hadrons, like pions, charged particles are selected if the observed $dE/dx$ is consistent with the expected values for a particular particle. A normalized $dE/dx$ is quantified as: $n_\sigma(\pi) = \frac{1}{\sigma_{exp}}\mathrm{ln}\left(\frac{dE/dx_{obs}}{dE/dx_{\pi,calc}}\right)$. Here $dE/dx_{obs}$ is the observed energy loss of the tracks in the TPC, $dE/dx_{\pi,calc}$ is the expected energy loss for charged pions based on the Bichsel formalism, and $\sigma_{exp}$ is the $dE/dx$ resolution of the TPC. Enhanced particle identification is facilitated by the Time of Flight (TOF) detector~\cite{STARTOF2005}, particularly useful when $n_\sigma(\pi)$ of two different particles are close. The mass square of a particle can be calculated by $m^{2} = p^{2}(1/\beta^{2}-1)$ with the momentum ($p$) measured from TPC and the inverse velocity $1/\beta$ from TOF. As can be seen from Fig.~\ref{fig:pid}, TOF provides very good separation of different particle species when their energy losses energy losses in TPC are close~\cite{STARPID2006}~\cite{STARPID2010}.\par

Jets are reconstructed using the anti-$k_{T}$ algorithm \cite{antikt}~\cite{fastjet} with the radius $R$ = 0.6 for the measurement at $\sqrt{s} =$ 200 GeV~\cite{STARjet2009pp200}. The off-axis cone method~\cite{STARjetdijet2012pp500} is adopted to correct for the underlying event contribution in the analysis. Both the jet energies and spin asymmetries are corrected for the smearing from the underlying event contamination.\par

\section{Results}
\begin{figure*}
\centering
\includegraphics[width=1.0\linewidth]{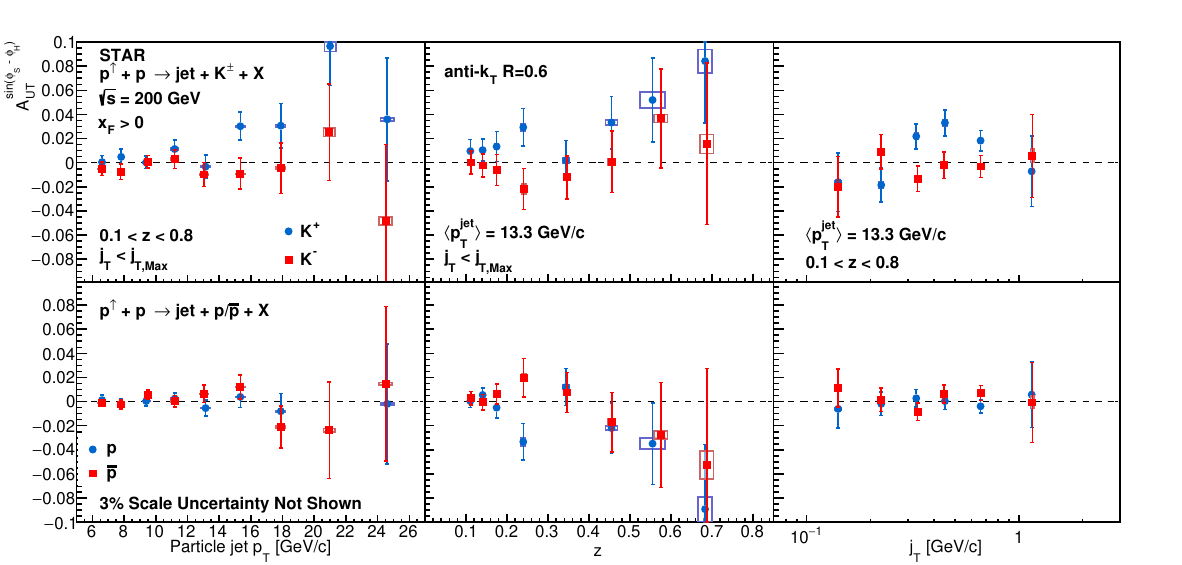}
\caption{Collins asymmetries, $A_{UT}^{\sin(\phi_{S}-\phi_{H})}$, as a function of particle jet-$p_{T}$, hadron-$z$, and hadron-$j_{T}$ for charged kaons (upper panels) and protons (lower panels) inside jets.}
\label{fig-2}       
\end{figure*}

Figure~\ref{fig-2} presents the first measurement of the Collins asymmetries for charged kaons (top panels) and protons (bottom panels) within jets in $pp$ collisions at $\sqrt{s} =$ 200~GeV~\cite{STAR:2022hqg}. These results are plotted as a function of jet transverse momentum ($p_{T}$), hadron momentum fractions ($z$), and hadron transverse momentum relative to the jet axis ($j_{T}$), arranged from left to right across the panels.\par 

Despite the limitations due to the statistics, results for $K^{+}$ exhibit positive asymmetries that align closely in magnitude with those observed for $\pi^{+}$~\cite{STAR:2022hqg}. In contrast, the asymmetries for \(K^{-}\) and protons are statistically consistent with zero at the one-sigma level, suggesting that the Collins effect is predominantly influenced by the favored fragmentation of valence quarks and is substantially reduced in unfavored fragmentation channels. As $K^{-}$ also includes an $s$ quark, this also indicates that its transversity should be much smaller than that of $u$ or $d$ quarks. Fragmentation into protons is much more complicated than into mesons, and the results here suggest that this process do not produce Collins asymmetries.\par

Additionally, recent results from STAR provided enriched measurements on elucidating the complex spin structures and dynamics in polarized proton-proton collisions. The measurement of the inclusive and diffractive electromagnetic jets at forward rapidity provide a deeper look into the origin of the large transverse single spin asymmetries~\cite{Liang:2023bmk}. While the spin-spin correlations between Lambda hyperons and Lambda hyperons in jet also explore the transverse polarization of $s$ quark inside the proton~\cite{Jan_spin2023}~\cite{Gao:2024dxl}.\par
\section{Summary}

The spin physics results from STAR have made significant progress in advancing our understanding of the internal spin structure of nucleons~\cite{GrossePerdekamp:2015xdx}~\cite{Achenbach:2023pba}. The venture into transverse spin measurements has produced important findings, and more theoretical efforts are needed to understand these new phenomenons. With its unique kinematic coverage, new and large data samples with both mid-rapidity and forward upgraded STAR detector will further elucidate TMD physics and unravel the complexities of nucleon spin dynamics before the arrival of Electron Ion Collider (EIC).\par

\section*{Acknowledgments}
This work was supported in part by the National Natural Science Foundation of China under Grant No.12375139 and Shandong Province Science Foundation for Youths under Grant No.ZR2022QA032.\par

\end{document}